\documentclass[prd,aps,preprint,groupedaddress,superscriptaddress,nofootinbib]{revtex4}
\usepackage{graphicx}
\usepackage{amsmath,amssymb,bm,mathrsfs}
\usepackage{hyperref}
\usepackage{autobreak,mathtools}
\usepackage[dvipsnames, usenames]{xcolor}
\usepackage[normalem]{ulem}
\usepackage{comment}

\definecolor{rRGB}{RGB}{171, 40, 52}
\hypersetup{colorlinks=true, citecolor=rRGB, linkcolor=rRGB,urlcolor=rRGB}

\definecolor{p-r}{RGB}{171, 40, 52}
\hypersetup{colorlinks=true, citecolor=p-r, linkcolor=p-r,urlcolor=p-r}

\begin{document}

\preprint{YITP-25-68, IPMU25-0020}

\title{Imprints of quantum vacuum fluctuations on the gravitational field of a spherical mass}
\author{Ra\'ul Carballo-Rubio}
\affiliation{Instituto de Astrof\'isica de Andaluc\'ia (IAA-CSIC),
Glorieta de la Astronom\'ia, 18008 Granada, Spain}
\affiliation{Center of Gravity, Niels Bohr Institute, Blegdamsvej 17, 2100 Copenhagen, Denmark}
\author{Francesco Di Filippo}
\affiliation{Institut f\"ur Theoretische Physik, Max-von-Laue-Str. 1, 60438 Frankfurt, Germany}
\author{Shinji Mukohyama}
\affiliation{Center for Gravitational Physics and Quantum Information, Yukawa Institute for Theoretical Physics, Kyoto University, Kyoto 606-8502, Japan}
\affiliation{Research Center for the Early Universe (RESCEU), Graduate School of Science, The University of Tokyo, Hongo 7-3-1, Bunkyo-ku, Tokyo 113-0033, Japan}
\affiliation{Kavli Institute for the Physics and Mathematics of the Universe (WPI),
The University of Tokyo Institutes for Advanced Study,
The University of Tokyo, Kashiwa, Chiba 277-8583, Japan}
\author{Kazumasa Okabayashi}
\affiliation{Center for Gravitational Physics and Quantum Information, Yukawa Institute for Theoretical Physics, Kyoto University, Kyoto 606-8502, Japan}

\begin{abstract}
The Schwarzschild geometry, describing the gravitational field of a spherical mass in classical vacuum, is one of the most famous vacuum solutions of the Einstein field equations. Classical vacuum is an idealization that does not include quantum vacuum fluctuations of quantum fields, and determining the form of the gravitational field of a spherical mass in quantum vacuum is an important step towards understanding the interplay between gravity and quantum field theory. We formulate and prove general results on the space of static, spherically symmetric and asymptotically flat spacetimes sourced by quantum vacuum fluctuations, obtained under the broad assumptions that the quantum vacuum energy density is negative and unbounded on Killing horizons. In particular, we show the generic replacement of Killing horizons by wormhole throats. We discuss how previous calculations in the literature that have used different prescriptions for the regularized vacuum expectation value of the quantum stress-energy tensor are particular cases of our general results. 
\end{abstract}

\maketitle

\section{Introduction}

General relativity is by now a century-old and successful theory~\cite{Ashtekar:2014bja}. One of its most famous solutions, the Schwarzschild metric, was found shortly after the publication of the Einstein field equations~\cite{Schwarzschild:1916uq}. The discovery of the Schwarzschild metric was essential for the success of general relativity, as most of the early predictions of the new theory were formulated in situations in which the Schwarzschild metric provides a good description~\cite{Will:2014kxa}. This solution has continued to play an important role in physics education and new research developments up to date, and it is one of the most recognizable concepts in modern theoretical physics.

Both general relativity and the Schwarzschild metric are expected to be approximate descriptions of nature. Unifying general relativity with quantum field theory, the second pillar of modern theoretical physics, has proven to be an extremely difficult challenge~\cite{Carlip:2015asa,Giddings:2022jda}. How the space of solutions of general relativity, and the Schwarzschild solution in particular, are affected by this unification is currently unknown. This is the problem partially addressed in this work.

The Schwarzschild metric describes the gravitational field around a spherical mass in classical vacuum. The notion of classical vacuum assumes that spacetime is completely devoid of matter and energy. However, quantum field theory clearly indicates that vacuum is more complex and that it actually has a rich structure. The notion of quantum vacuum comprises fluctuations of mass and energy that cancel on average in the absence of gravitational fields, thus macroscopically resembling classical vacuum (in technical terms, the renormalized expectation value of the stress-energy tensor operator of quantum fields vanishes). 

\begin{figure}[htbp]
\includegraphics[width=0.5\linewidth]{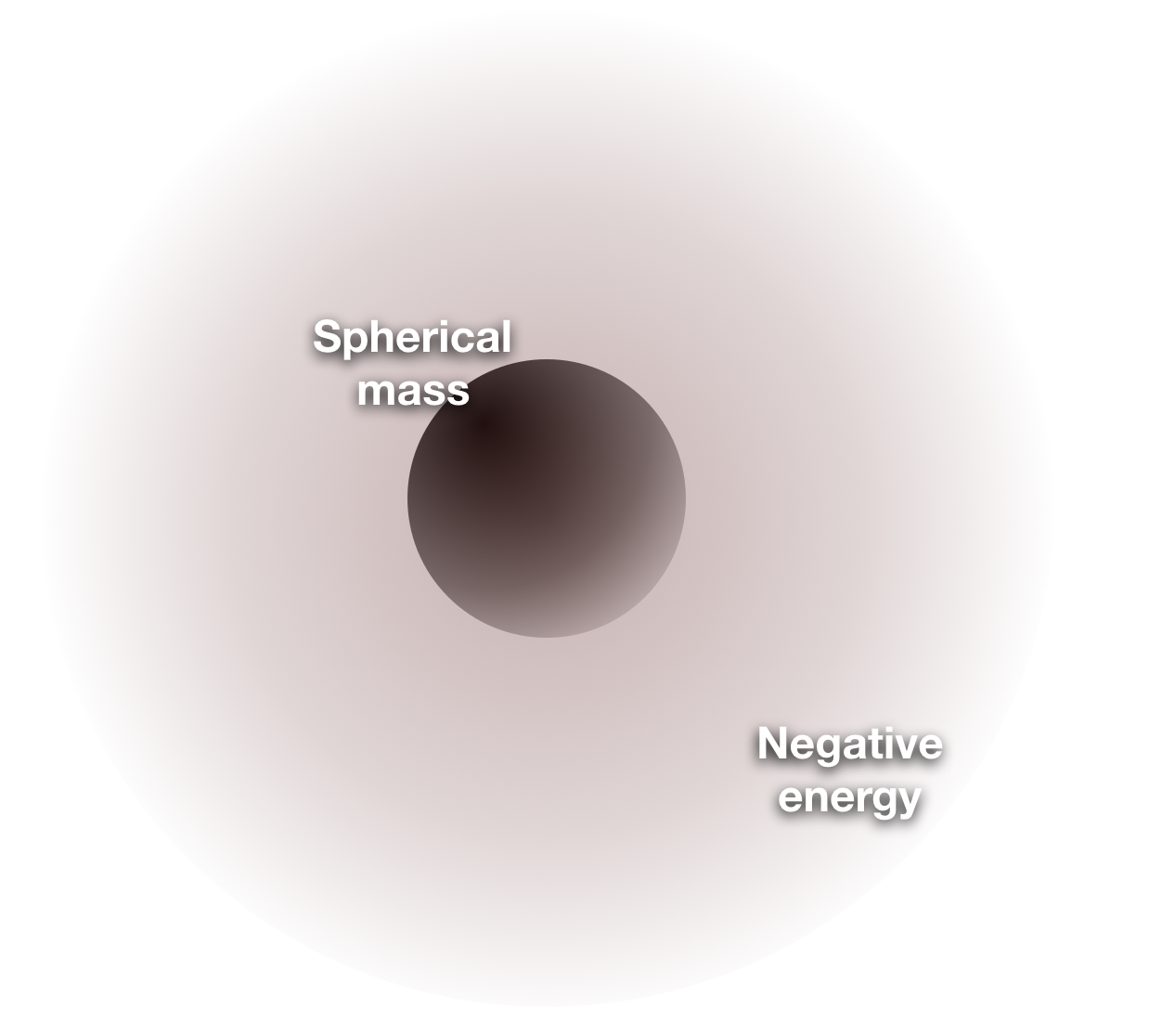}
\caption{\label{fig:polarization}
A spherical mass in quantum vacuum induces quantum vacuum polarization in its exterior, generating a cloud of negative energy density that changes the structure of the Schwarzschild metric. In this paper, we prove general results characterizing the spacetime generated by this cloud.}
\end{figure}

When a gravitational field due to a distribution of matter is present, quantum vacuum behaves differently than classical vacuum. This is due to the phenomenon of quantum vacuum polarization~\cite{Candelas:1980pca,Visser:1997gf}, first described for the electromagnetic field~\cite{Schwinger:1951nm}. The gravitational field breaks the balance between positive and negative fluctuations of mass and energy in the quantum vacuum, producing an energy deficit and an associated macroscopic gravitational field~\cite{Barcelo:2009tpa,Carballo-Rubio:2017tlh}. Hence, when the notion of classical vacuum is replaced by that of quantum vacuum, the Schwarzschild metric is expected to be replaced by a different metric structure (see Fig.~\ref{fig:polarization}). Characterizing these modifications is the goal of this work.

\section{Setup}

For a static, spherically symmetric and asymptotically flat spacetime, there always exist coordinates covering the asymptotic region in which the metric can be written as
\begin{equation}\label{eq:ansatz}
\text{d}s^2=-A(r)\text{d}t^2+\frac{\text{d}r^2}{B(r)}+r^2\text{d}\theta^2+r^2\sin^2\theta\text{d}\phi^2,    
\end{equation}
where $1-A(r)$ and $1-B(r)$ are both $\mathcal{O}(1/r)$ in the asymptotic region $r\rightarrow\infty$. This metric has a timelike Killing vector $\bm{\partial}_t$. 

These coordinates do not need to cover the whole manifold. The functions $A(r)$ and $B(r)$ are positive in the domain of validity of these coordinates, but can vanish in the limit in which its boundary is approached as we will discuss in more detail below. Another possibility is introducing horizon-penetrating coordinates instead.

The Schwarzschild solution arises as the solution to the vacuum Einstein field equations
\begin{equation}
\bm{G}=0,
\end{equation}
where $\bm{G}$ is the Einstein tensor. These equations constrain the functions $A(r)$ and $B(r)$ to take the form $A(r)=1/B(r)=1-2M/r$. Note that we will be using geometrized units in which $c=G=1$~\cite{Misner:1973prb}.

Our goal here is understanding how the Schwarzschild solution is modified when the right-hand side in the Einstein field equations is the expectation value of the stress-energy tensor operator $\hat{\bm{T}}$ of quantum fields in their vacuum state,
\begin{equation}\label{eq:seinf}
    \bm{G}=8\pi\langle 0|\hat{\bm{T}}|0\rangle.
\end{equation}
The expectation value $\langle 0|\hat{\bm{T}}|0\rangle$ is a functional of the metric that needs to be regularized. Due to the symmetries of the problem, the vacuum expectation value can always be written~\cite{Visser:1996iw,Visser:1996iv,Visser:1997sd} as a static and spherically symmetric anisotropic fluid (see~\cite{Cadogan:2024mcl} for a definition):
\begin{equation}\label{eq:anfl}
\langle 0|\hat{\bm{T}}|0\rangle=\rho\, \bm{u}\otimes\bm{u}+p_r\bm{r}\otimes\bm{r}+p_t\left(\bm{g}+\bm{u}\otimes\bm{u}-\bm{r}\otimes\bm{r}\right), 
\end{equation}
where $\bm{u}=A(r)^{-1/2}\bm{\partial}_t$ is a unit timelike vector representing the velocity field of the fluid and $\bm{r}=B(r)^{1/2}\bm{\partial}_r$ is a unit spacelike vector signaling the direction of anisotropy (the radial direction in this case).

Constructing closed expressions for $\langle 0|\hat{\bm{T}}|0\rangle$ or, equivalently, the three scalars $\rho$ (density), $p_r$ (radial pressure) and $p_t$ (tangential pressure) in terms of the metric functions $A(r)$ and $B(r)$ is an intricate problem that has been solved only for specific cases and approximations. We will briefly review some of the existing literature when discussing our results below.

However, we will bypass these difficulties by focusing on two aspects that are shared by different calculations. In particular, in \emph{every} situation studied to date, for static,
spherically-symmetric and asymptotically flat semiclassical
solutions, the quantum vacuum energy density is \emph{(i)} non-positive everywhere and \emph{(ii)} unbounded at the outermost positive root of $A(r)$ (i.e., the would-be Killing horizon of $\bm{\partial}_t$), if the latter exists. We take these two aspects as \emph{axiomatic} defining characteristics of the energy density associated with quantum vacuum fluctuations,\footnote{From the arguments below, it is obvious that we need to make these assumptions only in the region covered by the coordinates used in Eq.~\eqref{eq:ansatz}.} and focus on their implications. We will discuss evidence in favor of the plausibility of these assumptions in section
\ref{sec:discussion} when reviewing our results in the light of previous literature.

\section{Space of solutions}

We now formulate and prove a series of results for the spacetimes and sources discussed above.

\textbf{Proposition 1.} The metric function $A(r)$ must remain positive for the spacetime to be regular.

\textbf{Proof.} We construct a proof by contradiction. We first note that $A(r)$ at spatial infinity is positive. If $A(r)$ has a positive root then by assumption the energy density is unbounded at the outermost positive root $r=r_\star$ of $A(r)$:
\begin{equation}
\lim_{r\rightarrow r_\star}\rho(r)=-\infty.
\end{equation}
The Kretschmann scalar $K$ of the metric in Eq.~\eqref{eq:ansatz} is given by
\begin{equation}
K=\frac{1}{2}\sum_{n=1}^4 (K_n)^2,
\label{eq:Kretschmann}
\end{equation}
where $K_n$ are functions of the metric (see Appendix~\ref{sec:app}). Therefore, each contribution $(K_n)^2/2$ to the Kretschmann scalar is non-negative. Two of the terms are
\begin{equation}
 K_1=\frac{2}{r}B'=\frac{2}{r}\left( \frac{2m}{r^2}-\frac{2m'}{r}\right)=\frac{2}{r}\left( \frac{1-B}{r}-\frac{2m'}{r}\right),
\end{equation}
and
\begin{equation}
K_2=\frac{2\sqrt{2}\left(1-B\right)}{r^2},
\end{equation}
where $m'=\text{d}m/\text{d}r$ is the derivative of the Misner--Sharp (MS) mass~\cite{Misner:1964je,Hayward:1994bu}, $m=r(1-B)/2$, with respect to the area radius $r$. 
The temporal component of Eqs.~(\ref{eq:seinf}-\ref{eq:anfl}) can be written as
\begin{equation}\label{eq:derivative_of_MS}
 m'=4\pi r^2\rho.
\end{equation}
From $K_2$ we deduce that $B(r_\star)$ must be finite, and therefore
\begin{equation}
 \lim_{r\rightarrow r_\star}K_1=\frac{2}{r_\star^2}\left[1-B(r_\star)-\left.8\pi r_\star^2\rho(r)\right|_{r=r_\star} \right],
\end{equation}
must be unbounded, thus leading to a curvature singularity. Hence, it is not possible for $A(r)$ to have a positive root.\hfill$\square$

\textbf{Proposition 2.} For spacetimes with positive and finite Arnowitt--Deser--Misner (ADM) mass $M$, the function $B(r)$ in Eq.~\eqref{eq:ansatz} has at least one positive root and the coordinates used in Eq.~\eqref{eq:ansatz} cover the region $r\in(r_0,\infty]$ with $r_0\geq 2M$ being the outermost root of $B(r)$, provided the spacetime is regular in that region.

\textbf{Proof.} The MS mass reduces to the ADM mass $M$ for $r\rightarrow\infty$, and its derivative is given as Eq. \eqref{eq:derivative_of_MS}. Due to the assumption $\rho\leq 0$, it follows that $m(r)\geq M$ and
\begin{equation}\label{eq:prop2ineq}
\frac{2m(r)}{r}\geq \frac{2M}{r}.
\end{equation}
On the other hand, due to $A(r)>0$ implied by the regularity assumption and Proposition 1, the coordinates in Eq.~\eqref{eq:ansatz} remain valid whenever $B(r)>0$. If we further assume that $B(r)$ does not vanish for $r\geq 2M$ then it contradicts with Eq.~\eqref{eq:prop2ineq} since $B=1-2m/r$. Hence, there is a value $r_0\geq 2M$ for which
\begin{equation}
 B(r)>0\ \mbox{  for  }\ r>r_0, \quad \lim_{r\rightarrow r_0}B(r)=0.
\end{equation}
Note that $r_0\neq\infty$ due to the constraint that $m(\infty)=M$.\hfill$\square$

\textbf{Proposition 3.} For $r_0$ defined in Proposition 2, the hypersurface $r=r_0$, if regular, is a wormhole throat.

\textbf{Proof.} We can define the coordinate $x$ as
\begin{equation}\label{eq:wormhcoord}
\text{d}x=\frac{\text{d}r}{\sqrt{B(r)}},   
\end{equation}
in which the metric takes the form
\begin{equation}
\text{d}s^2=-A[r(x)]\text{d}t^2+\text{d}x^2+r(x)^2\left(\text{d}\theta^2+\sin^2\theta\text{d}\phi^2\right).    
\end{equation}
This metric is defined for $x\in(-\infty,\infty)$. From Eq.~\eqref{eq:wormhcoord}, the direct calculation of $\left.\text{d}^2r/\text{d}x^2\right|_{x=x_0}$ at $x_0=x(r_0)$ shows that
\begin{equation}
\left.\frac{\text{d}^2r}{\text{d}x^2}\right|_{x=x_0}=\frac{1}{2}\left.\frac{\text{d}B}{\text{d}r}\right|_{x=x_0}=\frac{1}{2r_0}\left[1-8\pi r_0^2\rho(r_0)\right]>0,
\end{equation}
due to the assumption $\rho\leq0$, thus finishing the proof~\cite{Visser:1997yn}.\hfill$\square$

\vspace{0.5cm}
\noindent
So far we have worked with Eqs.~\eqref{eq:seinf} in which the only source is originated from fluctuations of the quantum vacuum. However, for the next two propositions we introduce an additional contribution to the right hand side of Eqs.~\eqref{eq:seinf} from a fluid with a non-negative energy density $\rho_m\geq0$. We will still assume that the energy density of the vacuum remains non-positive, $\rho\leq 0$, in the presence of this additional matter source.

\textbf{Proposition 4.} 
Proposition 1 remains valid if $\rho_m$ is finite, while propositions 2 and 3 remain valid if $\rho_m+\rho\leq 0$ in the subset of the domain of dependence of $r$ in which $B(r)\geq0$.

\textbf{Proof.} All the steps in the proofs of these three propositions can be generalized to the situation with non-zero $\rho_m$ by replacing $\rho\rightarrow\rho_m+\rho$. The key behavior of the total energy density in the proof of Proposition 1 is that $\rho_m+\rho$ is divergent at $r_\star$, which is guaranteed if $\rho_m$ is finite. On the other hand, the key behavior of the energy density in Propositions 2 and 3 is that $\rho_m+\rho$ is non-positive for $r\geq r_0$.\hfill$\square$

\vspace{0.5cm}
\noindent
The proposition above illustrates how the existence of a wormhole throat is a feature of vacuum (without classical matter) or near-vacuum spacetimes. Including classical matter can result into black hole mimickers (see~\cite{Carballo-Rubio:2025fnc} for a definition and examples) without wormhole throats. This is also necessary to ensure that the topology of the spacetime is $\mathbb{R}^4$ in the presence of wormhole throats, as discussed in the propositions below.

\textbf{Proposition 5.} Suppose that $x=x_0$ ($>0$) is a regular wormhole throat. The topology of the spacetime cannot be $\mathbb{R}^4$ unless a spherical mass is present such that the total energy density is positive, and $\rho_m(r)>|\rho(r)|$, for some of its points beyond the wormhole throat, $x<x_0$.

\textbf{Proof.} By contradiction: if the topology is $\mathbb{R}^4$ then the function $r(x)$ must vanish at a value $\bar{x}<x_0$, while the mass needs to vanish at $r=0$ to avoid a conical singularity. However, we know that 
\begin{equation}
    r(x_0)>0\,,\qquad \left.\frac{\text{d}r}{\text{d}x}\right|_{x=x_0}=0\,,\qquad\left.\frac{\text{d}^2r}{\text{d}x^2}\right|_{x=x_0}>0\,.
\end{equation}
Given that the first derivative vanishes, while the second derivative is positive, for a sufficiently small and positive $\epsilon$ we have
\begin{equation}
    r(x_0-\epsilon)>0\,,\qquad \left.\frac{\text{d}r}{\text{d}x}\right|_{x=x_0-\epsilon}<0\,,\qquad\left.\frac{\text{d}^2r}{\text{d}x^2}\right|_{x=x_0}>0\,.
\end{equation}
On the other hand, at $\bar{x}$
\begin{equation}
    r(\bar{x})=0\,,\qquad \left.\frac{\text{d}r}{\text{d}x}\right|_{x=\bar{x}} = 1\,.
\end{equation}
Therefore, there has to be an extra zero of the derivative $\text{d}r/\text{d}x$ in the region $\bar{x}<x<x_0$, which is only possible if $\text{d}^2r/\text{d}x^2$ changes sign.

To complete the proof, we just need to observe that
\begin{equation}
\frac{\text{d}^2r}{\text{d}x^2}=\frac{1}{2r}\left\{1-B(r)-8\pi r^2\left[\rho(r)+\rho_m(r)\right]\right\},
\end{equation}
where we have added a spherical mass of the fluid with energy density $\rho_m(r)$. The left-hand side in the former equation changing sign requires that $\rho(r)+\rho_m(r)$ is positive in an open interval, and therefore that $\rho_m(r)>|\rho(r)|$ as well.\hfill$\square$

\textbf{Proposition 6.} The total positive energy of the spherical mass required to have $\mathbb{R}^4$ topology is bounded from below by a constant depending only on the total negative present in its exterior.

\textbf{Proof.} The ADM mass being positive, $M>0$, implies the inequality
\begin{equation}
\int_{\bar{x}}^\infty\text{d}x\, \frac{\text{d}}{\text{d}x} m(x) > 0, 
\end{equation}
which, taking into account the chain rule, implies
\begin{equation}
\int_{\bar{x}}^\infty\text{d}x\,r^2(x)\frac{\text{d}r(x)}{\text{d}x}\left[\rho_m(x)+\rho(x)\right]>0.  
\end{equation}
We can split the real line into intervals according to the sign of $\text{d}r/\text{d}x$, so that the latter is positive in the intervals $\{U_k=(x^-_k,x^+_k)\}_{k=1}^M$, negative for $\{V_l=(x^-_l,x^+_l)\}_{l=1}^N$, and vanishes at the boundaries of these intervals. The union of all these intervals is $[\bar{x},+\infty)$, thus we can write
\begin{equation}
\sum_{k=1}^M\int_{U_k}\text{d}x\,r^2(x)\left|\frac{\text{d}r(x)}{\text{d}x}\right|\left[\rho_m(x)+\rho(x)\right]-\sum_{l=1}^N\int_{V_l}\text{d}x\,r^2(x)\left|\frac{\text{d}r(x)}{\text{d}x}\right|\left[\rho_m(x)+\rho(x)\right]>0.  
\end{equation}
As the function $r(x)$ is single-valued in the domain of integration of each of the integrals appearing in the expression above, we can use instead the variable $r$ and write
\begin{equation}
\sum_{k=1}^M\int_{r(x^-_k)}^{r(x^+_k)}\text{d}r\,r^2\left[\rho_m(r)+\rho(r)\right]+\sum_{l=1}^N\int_{r(x^+_l)}^{r(x^-_l)}\text{d}r\,r^2\left[\rho_m(r)+\rho(r)\right]>0, 
\end{equation}
where we have also inverted the limits of integration of the integrals over the intervals $\{V_l\}_{l=1}^N$, for which the set of relations $\{r(x^-_l)> r(x^+_l)\}_{l=1}^N$ are satisfied.

The final step in our proof is writing
\begin{equation}
\sum_{k=1}^M\int_{r(x^-_k)}^{r(x^+_k)}\text{d}r\,r^2\rho_m(r)+\sum_{l=1}^N\int_{r(x^+_l)}^{r(x^-_l)}\text{d}r\,r^2\rho_m(r)>-\sum_{k=1}^M\int_{r(x^-_k)}^{r(x^+_k)}\text{d}r\,r^2\rho(r)-\sum_{l=1}^N\int_{r(x^+_l)}^{r(x^-_l)}\text{d}r\,r^2\rho(r), 
\end{equation}
and noticing that the integrals on the right-hand side are greater or equal to the quantity resulting from restricting the integration to the interval to $x\in [x_s,+\infty)$, where $R=R(x_s)$ is the radius of the surface of the classical matter content.
\hfill$\square$

\section{Discussion}\label{sec:discussion}

The calculation of the renormalized stress-energy tensor, which is an essential piece in the semiclassical Einstein field equations, is a complex undertaking plagued by renormalization and calculational ambiguities. Previous works have dealt with these ambiguities by imposing assumptions that generally limit their scope of applicability. In this work we propose a different approach, in which general results are proven without committing to a particular approximation scheme, but are rather based in generic features shared by different approximations. In this section, we discuss the interplay between previous works and our results.

The generic features at the core of our results are the assumptions that the energy density generated by gravitational vacuum polarization is non-positive between the asymptotically flat region and the surface of infinite blueshift 
(equivalently, the would-be Killing horizon of $\bm{\partial}_t)$ and unbounded on the latter: an
observer sitting arbitrarily close to the hypersurface defined by the outermost positive root of $A(r)$ would detect
photons coming from outside that are highly blueshifted. The local frequencies of mode functions that enter into the calculation of the renormalized stress-energy tensor are also blueshifted around this hypersurface, leading to the divergent behavior of the latter.

\begin{figure}[htbp]
\includegraphics[width=0.6\linewidth]{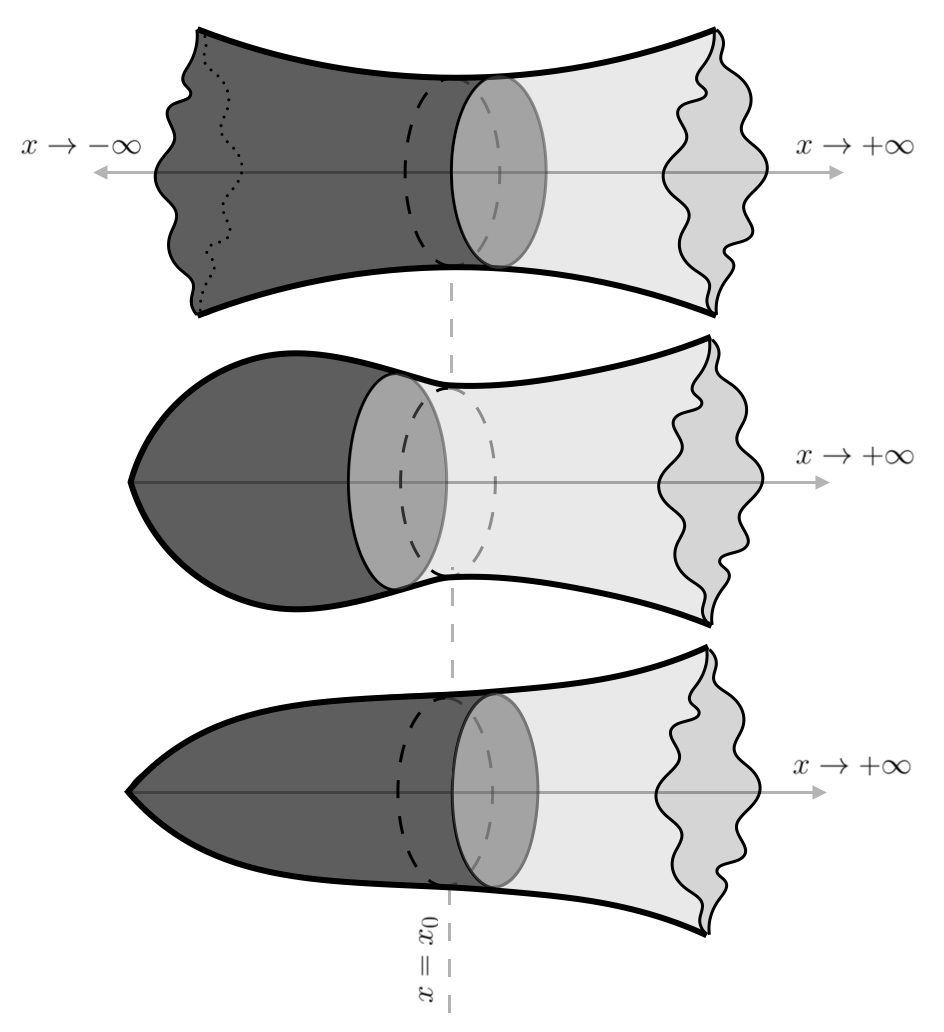}
\caption{Graphical representation of $r(x)$ for three different spacetimes. The dark (light) gray region indicates the spacetime inside (outside) a spherical mass with radius $R$, which can be greater or lower than $r(x_0)$. The topology of the resulting spacetime depends on the relative value of the density $\rho_m(r)$ when compared to the energy density associated with quantum vacuum fluctuations $\rho(r)$. Lower values of $R$ generally require greater values of $\rho_m(r)$ in order to maintain a $\mathbb{R}^4$ topology. Additional details for solutions of this kind across different approximations can be found in~\cite{Ho:2017vgi,Arrechea:2021pvg,Arrechea:2021xkp,Arrechea:2023oax,Arrechea:2023wgy}.
\label{fig:spacetimes}}
\end{figure}

Under these assumptions, some qualitative properties of the spacetime outside a spherical mass surrounded by quantum vacuum can be extracted. It is convenient to describe it resorting to a thought experiment in which we start with a spherical mass with radius $R\gg 2M$ and decrease this radius gradually, uncovering the features of the spacetime around it. Assuming that the spacetime remains regular, static and asymptotically flat as the radius $R$ of the spherical mass decreases, the spacetime metric shows deviations from the Schwarzschild metric until eventually developing a wormhole throat at $r=r_0$. Whenever a wormhole throat is developed, the topology of the resulting spacetime may be determined by the amount of positive energy matter in the spacetime, leading to different possibilities as depicted in Fig.~\ref{fig:spacetimes}.

The motivation for our analysis comes from the numerous previous works finding a wormhole structure in different approximations (see~\cite{Ho:2017joh,Ho:2017vgi,Berthiere:2017tms,Ho:2018jkm,Fabbri:2005zn,Arrechea:2019jgx,Arrechea:2021ldl,Arrechea:2022dvy,Beltran-Palau:2022nec,Kain:2025ygm,Kain:2025lxd} for a non-exhaustive list), which pointed towards some kind of universality. We have shown above that the existence of wormhole throats is indeed associated with a very specific behavior of the energy density associated with quantum vacuum fluctuations, namely that the latter is non-positive and unbounded on surfaces of infinite blueshift. All the works mentioned above display these features, in accordance to our results, while also showing other properties such as the occurrence of singularities (e.g., on the other side of the wormhole throat) that we have not discussed.

Both features are characteristic of the Boulware vacuum~\cite{Boulware:1974dm,Boulware:1975fe} when evaluated on the Schwarzschild background either using numerical or analytical approximations~\cite{Candelas:1980pca,Anderson:1994hg,Visser:1996iv,Mottola:2006ew,Arrechea:2024cnv}. Indeed, the renormalized stress-energy tensor for the Hartle-Hawking state is essentially the same
as a “topped-up” Boulware state~\cite{Mukohyama:1998rf} whose
temperature is divergent on the would-be Killing horizon due to an
infinite blueshift. This, combined with the fact that the
renormalized stress-energy tensor for the Hartle-Hawking state remains finite, implies that the Boulware vacuum
energy density is negatively divergent at the would-be Killing
horizon. Note that the Hartle-Hawking state itself does not provide a counterexample to our assumptions since it does not lead to an asymptotically flat static solution when backreaction is taken into account. To the best of our knowledge, there is no general proof that the quantum vacuum energy density must be negative and unbounded on surfaces of infinite blueshift, although it is certainly reasonable and there are ample cumulative evidences that both conditions are satisfied. Cumulative evidences include all the works studying backreaction mentioned above, in which these properties are satisfied~\cite{Berthiere:2017tms,Fabbri:2005zn,Arrechea:2019jgx,Arrechea:2021ldl,Arrechea:2022dvy,Beltran-Palau:2022nec,Kain:2025ygm,Kain:2025lxd}. Strong additional evidence for the quantum vacuum energy density being unbounded on surfaces of infinite blueshift in the presence of backreaction comes from the analytical Polyakov approximation, in which this statement is strictly true as the quantum vacuum energy density in the Boulware vacuum is inversely proportional to the norm of $\bm{\partial}_t$ in any static spacetime~\cite{Barcelo:2011bb}.

Finally, let us stress that, in this work, we are only accounting for the effects of quantum vacuum fluctuations of matter fields. We have characterized the imprints of semiclassical modifications as appearing on the right-hand side of Eq.~\eqref{eq:seinf} while keeping the structure of the left-hand side of these equations untouched, which is a standard procedure in the study of semiclassical backreaction. Analyzing a more general problem using field equations that capture the most general second-order modifications of the left-hand side of the Einstein field equations~\cite{Carballo-Rubio:2025ntd}, recasting this more general problem in a Hamiltonian setting~\cite{Alonso-Bardaji:2025hda}, or including higher-curvature corrections, would generalize our results and might uncover other geometric possibilities.

\section{Conclusions}

We have shown that it is possible to derive generic features of the gravitational field of a spherical mass in quantum vacuum in a robust manner, from simple assumptions about the behavior of the negative energy density associated with gravitational vacuum polarization. The spacetimes obtained in this work supersede the Schwarzschild metric as a more accurate description of the static and asymptotically flat gravitational field outside a spherical mass, properly accounting for the microstructure of the quantum vacuum, and reducing to the former in the $\hbar\rightarrow 0$ limit.

Our results solidify existing studies that relied on assumptions of diverse nature to simplify the necessary calculations, showing the emergence of universal features and pinpointing the reasons for this universality. This study thus offers a proof of principle that renormalization and calculational ambiguities do not prevent the extraction of robust conclusions in a semiclassical framework. It also underlines the importance of using similar methods in more complex scenarios such as gravitational collapse, which could lead to robust results and possibly new insights about the nature of black holes.

\appendix

\section{Kretschmann scalar}\label{sec:app}

For any static spacetime, the Kretschmann scalar can be written as a sum of squares~\cite{Lobo:2020ffi}. The expression for the Kretschmann scalar for the metric \eqref{eq:ansatz} used in this paper is:
\begin{equation}
K=\frac{4}{r^4}\left\{r^2\frac{\left(B'\right)^2}{2}+r^2\frac{B^2\left(A'\right)^2}{2A^2}+\left(-1+B\right)^2+r^4\left[\frac{BA''}{2A}+\frac{A'B'}{4A}-\frac{B\left(A'\right)^2}{4A^2}\right]^2\right\},
\end{equation}
so that we can define
\begin{align}
K_1&=\frac{2B'}{r},\nonumber\\ 
K_2&=\frac{2\sqrt{2}\left(1-B\right)}{r^2},\nonumber\\
K_3&=\frac{2BA'}{rA},\nonumber\\
K_4&=2\sqrt{2}\left[\frac{BA''}{2A}+\frac{A'B'}{4A}-\frac{B\left(A'\right)^2}{4A^2}\right],
\end{align}
to recast the Kretschmann scalar in the form \eqref{eq:Kretschmann}. 

\begin{acknowledgments}
RCR acknowledges financial support provided by the Spanish Government through the Ram\'on y Cajal program (contract RYC2023-045894-I) and the Grant No.~PID2023-149018NB-C43 funded~by MCIN/AEI/10.13039/501100011033, and by the Junta de Andaluc\'{\i}a 
through the project FQM219 and from the Severo Ochoa grant 
CEX2021-001131-S funded by MCIN/AEI/ 10.13039/501100011033, as well as the hospitality of the Center of Gravity, a Center of Excellence funded by the Danish National Research Foundation under grant No.~184. The work of FDF is supported by the
Alexander von Humboldt foundation. The work of SM was supported in part by Japan Society for the Promotion of Science Grants-in-Aid for Scientific Research No.~24K07017 and the World Premier International Research Center Initiative (WPI), MEXT, Japan. This work of KO is supported in part by Japan Society for the Promotion of Science (JSPS) KAKENHI Grant Numbers JP23KJ1162.
\end{acknowledgments}

\bibliographystyle{utphys}

\bibliography{refs}

\providecommand{\href}[2]{#2}\begingroup\raggedright\begin{thebibliography}{10}

\bibitem{Ashtekar:2014bja}
A.~Ashtekar, B.~K. Berger, J.~Isenberg, and M.~A.~H. MacCallum, ``{General
  Relativity and Gravitation: A Centennial Perspective},''
  \href{http://arxiv.org/abs/1409.5823}{{\ttfamily arXiv:1409.5823 [gr-qc]}}.

\bibitem{Schwarzschild:1916uq}
K.~Schwarzschild, ``{On the gravitational field of a mass point according to
  Einstein's theory},'' {\em Sitzungsber. Preuss. Akad. Wiss. Berlin (Math.
  Phys. )} {\bfseries 1916} (1916) 189--196,
  \href{http://arxiv.org/abs/physics/9905030}{{\ttfamily
  arXiv:physics/9905030}}.

\bibitem{Will:2014kxa}
C.~M. Will, ``{The Confrontation between General Relativity and Experiment},''
  \href{http://dx.doi.org/10.12942/lrr-2014-4}{{\em Living Rev. Rel.}
  {\bfseries 17} (2014) 4}, \href{http://arxiv.org/abs/1403.7377}{{\ttfamily
  arXiv:1403.7377 [gr-qc]}}.

\bibitem{Carlip:2015asa}
S.~Carlip, D.-W. Chiou, W.-T. Ni, and R.~Woodard, ``{Quantum Gravity: A Brief
  History of Ideas and Some Prospects},''
  \href{http://dx.doi.org/10.1142/S0218271815300281}{{\em Int. J. Mod. Phys. D}
  {\bfseries 24} no.~11, (2015) 1530028},
  \href{http://arxiv.org/abs/1507.08194}{{\ttfamily arXiv:1507.08194 [gr-qc]}}.

\bibitem{Giddings:2022jda}
S.~B. Giddings, ``{The deepest problem: some perspectives on quantum
  gravity},'' \href{http://arxiv.org/abs/2202.08292}{{\ttfamily
  arXiv:2202.08292 [hep-th]}}.

\bibitem{Candelas:1980pca}
P.~{Candelas}, ``{Vacuum polarization in Schwarzschild spacetime},''
  \href{http://dx.doi.org/10.1103/PhysRevD.21.2185}{{\em \prd} {\bfseries 21}
  no.~8, (Apr., 1980) 2185--2202}.

\bibitem{Visser:1997gf}
M.~Visser, ``{Gravitational vacuum polarization},'' in {\em {8th Marcel
  Grossmann Meeting on Recent Developments in Theoretical and Experimental
  General Relativity, Gravitation and Relativistic Field Theories (MG 8)}},
  pp.~842--844.
\newblock 6, 1997.
\newblock \href{http://arxiv.org/abs/gr-qc/9710034}{{\ttfamily
  arXiv:gr-qc/9710034}}.

\bibitem{Schwinger:1951nm}
J.~S. Schwinger, ``{On gauge invariance and vacuum polarization},''
  \href{http://dx.doi.org/10.1103/PhysRev.82.664}{{\em Phys. Rev.} {\bfseries
  82} (1951) 664--679}.

\bibitem{Barcelo:2009tpa}
C.~Barcel\'o, S.~Liberati, S.~Sonego, and M.~Visser, ``{Black Stars, Not
  Holes},'' \href{http://dx.doi.org/10.1038/scientificamerican1009-38}{{\em
  Sci. Am.} {\bfseries 301} no.~4, (2009) 38--45}.

\bibitem{Carballo-Rubio:2017tlh}
R.~Carballo-Rubio, ``{Stellar equilibrium in semiclassical gravity},''
  \href{http://dx.doi.org/10.1103/PhysRevLett.120.061102}{{\em Phys. Rev.
  Lett.} {\bfseries 120} no.~6, (2018) 061102},
  \href{http://arxiv.org/abs/1706.05379}{{\ttfamily arXiv:1706.05379 [gr-qc]}}.

\bibitem{Misner:1973prb}
C.~W. Misner, K.~S. Thorne, and J.~A. Wheeler, {\em {Gravitation}}.
\newblock W. H. Freeman, San Francisco, 1973.

\bibitem{Visser:1996iw}
M.~Visser, ``{Gravitational vacuum polarization. 1: Energy conditions in the
  Hartle-Hawking vacuum},''
  \href{http://dx.doi.org/10.1103/PhysRevD.54.5103}{{\em Phys. Rev. D}
  {\bfseries 54} (1996) 5103--5115},
  \href{http://arxiv.org/abs/gr-qc/9604007}{{\ttfamily arXiv:gr-qc/9604007}}.

\bibitem{Visser:1996iv}
M.~Visser, ``{Gravitational vacuum polarization. 2: Energy conditions in the
  Boulware vacuum},'' \href{http://dx.doi.org/10.1103/PhysRevD.54.5116}{{\em
  Phys. Rev. D} {\bfseries 54} (1996) 5116--5122},
  \href{http://arxiv.org/abs/gr-qc/9604008}{{\ttfamily arXiv:gr-qc/9604008}}.

\bibitem{Visser:1997sd}
M.~Visser, ``{Gravitational vacuum polarization. 4: Energy conditions in the
  Unruh vacuum},'' \href{http://dx.doi.org/10.1103/PhysRevD.56.936}{{\em Phys.
  Rev. D} {\bfseries 56} (1997) 936--952},
  \href{http://arxiv.org/abs/gr-qc/9703001}{{\ttfamily arXiv:gr-qc/9703001}}.

\bibitem{Cadogan:2024mcl}
T.~Cadogan and E.~Poisson, ``{Self-gravitating anisotropic fluids. I: Context
  and overview},'' \href{http://arxiv.org/abs/2406.03185}{{\ttfamily
  arXiv:2406.03185 [gr-qc]}}.

\bibitem{Misner:1964je}
C.~W. Misner and D.~H. Sharp, ``{Relativistic equations for adiabatic,
  spherically symmetric gravitational collapse},''
  \href{http://dx.doi.org/10.1103/PhysRev.136.B571}{{\em Phys. Rev.} {\bfseries
  136} (1964) B571--B576}.

\bibitem{Hayward:1994bu}
S.~A. Hayward, ``{Gravitational energy in spherical symmetry},''
  \href{http://dx.doi.org/10.1103/PhysRevD.53.1938}{{\em Phys. Rev. D}
  {\bfseries 53} (1996) 1938--1949},
  \href{http://arxiv.org/abs/gr-qc/9408002}{{\ttfamily arXiv:gr-qc/9408002}}.

\bibitem{Visser:1997yn}
M.~Visser and D.~Hochberg, ``{Generic wormhole throats},'' {\em Annals Israel
  Phys. Soc.} {\bfseries 13} (1997) 249,
  \href{http://arxiv.org/abs/gr-qc/9710001}{{\ttfamily arXiv:gr-qc/9710001}}.

\bibitem{Carballo-Rubio:2025fnc}
R.~Carballo-Rubio {\em et~al.}, ``{Towards a non-singular paradigm of black
  hole physics},'' \href{http://dx.doi.org/10.1088/1475-7516/2025/05/003}{{\em
  JCAP} {\bfseries 05} (2025) 003},
  \href{http://arxiv.org/abs/2501.05505}{{\ttfamily arXiv:2501.05505 [gr-qc]}}.

\bibitem{Ho:2017vgi}
P.-M. Ho and Y.~Matsuo, ``{Static Black Hole and Vacuum Energy: Thin Shell and
  Incompressible Fluid},''
  \href{http://dx.doi.org/10.1007/JHEP03(2018)096}{{\em JHEP} {\bfseries 03}
  (2018) 096}, \href{http://arxiv.org/abs/1710.10390}{{\ttfamily
  arXiv:1710.10390 [hep-th]}}.

\bibitem{Arrechea:2021pvg}
J.~Arrechea, C.~Barcel{\'o}, R.~Carballo-Rubio, and L.~J. Garay,
  ``{Semiclassical constant-density spheres in a regularized Polyakov
  approximation},'' \href{http://dx.doi.org/10.1103/PhysRevD.104.084071}{{\em
  Phys. Rev. D} {\bfseries 104} no.~8, (2021) 084071},
  \href{http://arxiv.org/abs/2105.11261}{{\ttfamily arXiv:2105.11261 [gr-qc]}}.

\bibitem{Arrechea:2021xkp}
J.~Arrechea, C.~Barcel{\'o}, R.~Carballo-Rubio, and L.~J. Garay,
  ``{Semiclassical relativistic stars},''
  \href{http://dx.doi.org/10.1038/s41598-022-19836-8}{{\em Sci. Rep.}
  {\bfseries 12} no.~1, (2022) 15958},
  \href{http://arxiv.org/abs/2110.15808}{{\ttfamily arXiv:2110.15808 [gr-qc]}}.

\bibitem{Arrechea:2023oax}
J.~Arrechea, C.~Barcel{\'o}, R.~Carballo-Rubio, and L.~J. Garay,
  ``{Ultracompact horizonless objects in order-reduced semiclassical
  gravity},'' \href{http://dx.doi.org/10.1103/PhysRevD.109.104056}{{\em Phys.
  Rev. D} {\bfseries 109} no.~10, (2024) 104056},
  \href{http://arxiv.org/abs/2310.12668}{{\ttfamily arXiv:2310.12668 [gr-qc]}}.

\bibitem{Arrechea:2023wgy}
J.~Arrechea, {\em {Hydrostatic equilibrium in the semiclassical
  approximation}}.
\newblock PhD thesis, Universidad de Granada, 2023.

\bibitem{Ho:2017joh}
P.-M. Ho and Y.~Matsuo, ``{Static Black Holes With Back Reaction From Vacuum
  Energy},'' \href{http://dx.doi.org/10.1088/1361-6382/aaac8f}{{\em Class.
  Quant. Grav.} {\bfseries 35} no.~6, (2018) 065012},
  \href{http://arxiv.org/abs/1703.08662}{{\ttfamily arXiv:1703.08662
  [hep-th]}}.

\bibitem{Berthiere:2017tms}
C.~Berthiere, D.~Sarkar, and S.~N. Solodukhin, ``{The fate of black hole
  horizons in semiclassical gravity},''
  \href{http://dx.doi.org/10.1016/j.physletb.2018.09.027}{{\em Phys. Lett. B}
  {\bfseries 786} (2018) 21--27},
  \href{http://arxiv.org/abs/1712.09914}{{\ttfamily arXiv:1712.09914
  [hep-th]}}.

\bibitem{Ho:2018jkm}
P.-M. Ho and Y.~Matsuo, ``{On the Near-Horizon Geometry of an Evaporating Black
  Hole},'' \href{http://dx.doi.org/10.1007/JHEP07(2018)047}{{\em JHEP}
  {\bfseries 07} (2018) 047}, \href{http://arxiv.org/abs/1804.04821}{{\ttfamily
  arXiv:1804.04821 [hep-th]}}.

\bibitem{Fabbri:2005zn}
A.~Fabbri, S.~Farese, J.~Navarro-Salas, G.~J. Olmo, and H.~Sanchis-Alepuz,
  ``{Semiclassical zero-temperature corrections to Schwarzschild spacetime and
  holography},'' \href{http://dx.doi.org/10.1103/PhysRevD.73.104023}{{\em Phys.
  Rev. D} {\bfseries 73} (2006) 104023},
  \href{http://arxiv.org/abs/hep-th/0512167}{{\ttfamily arXiv:hep-th/0512167}}.

\bibitem{Arrechea:2019jgx}
J.~Arrechea, C.~Barcel{\'o}, R.~Carballo-Rubio, and L.~J. Garay,
  ``{Schwarzschild geometry counterpart in semiclassical gravity},''
  \href{http://dx.doi.org/10.1103/PhysRevD.101.064059}{{\em Phys. Rev. D}
  {\bfseries 101} no.~6, (2020) 064059},
  \href{http://arxiv.org/abs/1911.03213}{{\ttfamily arXiv:1911.03213 [gr-qc]}}.

\bibitem{Arrechea:2021ldl}
J.~Arrechea, C.~Barcel{\'o}, R.~Carballo-Rubio, and L.~J. Garay,
  ``{Reissner{\textendash}Nordstr{\"o}m geometry counterpart in semiclassical
  gravity},'' \href{http://dx.doi.org/10.1088/1361-6382/abf628}{{\em Class.
  Quant. Grav.} {\bfseries 38} no.~11, (2021) 115014},
  \href{http://arxiv.org/abs/2102.03544}{{\ttfamily arXiv:2102.03544 [gr-qc]}}.

\bibitem{Arrechea:2022dvy}
J.~Arrechea, C.~Barcel{\'o}, R.~Carballo-Rubio, and L.~J. Garay,
  ``{Asymptotically flat vacuum solutions in order-reduced semiclassical
  gravity},'' \href{http://dx.doi.org/10.1103/PhysRevD.107.085005}{{\em Phys.
  Rev. D} {\bfseries 107} no.~8, (2023) 085005},
  \href{http://arxiv.org/abs/2212.09375}{{\ttfamily arXiv:2212.09375 [gr-qc]}}.

\bibitem{Beltran-Palau:2022nec}
P.~Beltr{\'a}n-Palau, A.~del R{\'\i}o, and J.~Navarro-Salas, ``{Quantum
  corrections to the Schwarzschild metric from vacuum polarization},''
  \href{http://dx.doi.org/10.1103/PhysRevD.107.085023}{{\em Phys. Rev. D}
  {\bfseries 107} no.~8, (2023) 085023},
  \href{http://arxiv.org/abs/2212.08089}{{\ttfamily arXiv:2212.08089 [gr-qc]}}.

\bibitem{Kain:2025ygm}
B.~Kain, ``{Quantum corrected Einstein-Yang-Mills black holes in semiclassical
  gravity},'' \href{http://dx.doi.org/10.1103/PhysRevD.111.044033}{{\em Phys.
  Rev. D} {\bfseries 111} no.~4, (2025) 044033},
  \href{http://arxiv.org/abs/2502.08278}{{\ttfamily arXiv:2502.08278 [gr-qc]}}.

\bibitem{Kain:2025lxd}
B.~Kain, ``{Dynamical evolution and stability of quantum corrected
  Schwarzschild black holes in semiclassical gravity},''
  \href{http://dx.doi.org/10.1103/ms6s-92tx}{{\em Phys. Rev. D} {\bfseries 111}
  no.~12, (2025) 124057}, \href{http://arxiv.org/abs/2507.00109}{{\ttfamily
  arXiv:2507.00109 [gr-qc]}}.

\bibitem{Boulware:1974dm}
D.~G. Boulware, ``{Quantum Field Theory in Schwarzschild and Rindler Spaces},''
  \href{http://dx.doi.org/10.1103/PhysRevD.11.1404}{{\em Phys. Rev. D}
  {\bfseries 11} (1975) 1404}.

\bibitem{Boulware:1975fe}
D.~G. Boulware, ``{Hawking Radiation and Thin Shells},''
  \href{http://dx.doi.org/10.1103/PhysRevD.13.2169}{{\em Phys. Rev. D}
  {\bfseries 13} (1976) 2169}.

\bibitem{Anderson:1994hg}
P.~R. Anderson, W.~A. Hiscock, and D.~A. Samuel, ``{Stress-energy tensor of
  quantized scalar fields in static spherically symmetric space-times},''
  \href{http://dx.doi.org/10.1103/PhysRevD.51.4337}{{\em Phys. Rev. D}
  {\bfseries 51} (1995) 4337--4358}.

\bibitem{Mottola:2006ew}
E.~Mottola and R.~Vaulin, ``{Macroscopic Effects of the Quantum Trace
  Anomaly},'' \href{http://dx.doi.org/10.1103/PhysRevD.74.064004}{{\em Phys.
  Rev. D} {\bfseries 74} (2006) 064004},
  \href{http://arxiv.org/abs/gr-qc/0604051}{{\ttfamily arXiv:gr-qc/0604051}}.

\bibitem{Arrechea:2024cnv}
J.~Arrechea, C.~Breen, A.~Ottewill, L.~Pisani, and P.~Taylor, ``{Renormalized
  stress-energy tensor for scalar fields in the Boulware state with
  applications to extremal black holes},''
  \href{http://dx.doi.org/10.1103/PhysRevD.111.085009}{{\em Phys. Rev. D}
  {\bfseries 111} no.~8, (2025) 085009},
  \href{http://arxiv.org/abs/2409.04528}{{\ttfamily arXiv:2409.04528 [gr-qc]}}.

\bibitem{Mukohyama:1998rf}
S.~Mukohyama and W.~Israel, ``{Black holes, brick walls and the Boulware
  state},'' \href{http://dx.doi.org/10.1103/PhysRevD.58.104005}{{\em Phys. Rev.
  D} {\bfseries 58} (1998) 104005},
  \href{http://arxiv.org/abs/gr-qc/9806012}{{\ttfamily arXiv:gr-qc/9806012}}.

\bibitem{Barcelo:2011bb}
C.~Barcelo, R.~Carballo, and L.~J. Garay, ``{Two formalisms, one renormalized
  stress-energy tensor},''
  \href{http://dx.doi.org/10.1103/PhysRevD.85.084001}{{\em Phys. Rev. D}
  {\bfseries 85} (2012) 084001},
  \href{http://arxiv.org/abs/1112.0489}{{\ttfamily arXiv:1112.0489 [gr-qc]}}.

\bibitem{Carballo-Rubio:2025ntd}
R.~Carballo-Rubio, ``{Master field equations for spherically symmetric
  gravitational fields beyond general relativity},''
  \href{http://dx.doi.org/10.1038/s41467-026-69035-6}{{\em Nature Commun.}
  {\bfseries 17} no.~1, (2026) 1399},
  \href{http://arxiv.org/abs/2507.15920}{{\ttfamily arXiv:2507.15920 [gr-qc]}}.

\bibitem{Alonso-Bardaji:2025hda}
A.~Alonso-Bardaji and D.~Brizuela, ``{Dynamical theory for spherical black
  holes in modified gravity},'' \href{http://dx.doi.org/10.1103/ttfy-yjdh}{{\em
  Phys. Rev. D} {\bfseries 112} no.~10, (2025) 104036},
  \href{http://arxiv.org/abs/2507.19380}{{\ttfamily arXiv:2507.19380 [gr-qc]}}.

\bibitem{Lobo:2020ffi}
F.~S.~N. Lobo, M.~E. Rodrigues, M.~V. de~Sousa~Silva, A.~Simpson, and
  M.~Visser, ``{Novel black-bounce spacetimes: wormholes, regularity, energy
  conditions, and causal structure},''
  \href{http://dx.doi.org/10.1103/PhysRevD.103.084052}{{\em Phys. Rev. D}
  {\bfseries 103} no.~8, (2021) 084052},
  \href{http://arxiv.org/abs/2009.12057}{{\ttfamily arXiv:2009.12057 [gr-qc]}}.

\end{thebibliography}\endgroup

\end{document}